# GEE-OPs: An Operator Knowledge Base for Geospatial Code Generation on the Google Earth Engine Platform Powered by Large Language Models

Shuyang Hou[a], Jianyuan Liang[a], Anqi Zhao[a], Huayi Wu[a]*

*a. State Key Laboratory of Information Engineering in Surveying, Mapping, and Remote Sensing, Wuhan University, Wuhan, China

*Corresponding author: Huayi Wu, email: <wuhuayi@whu.edu.cn>

**Abstract**

As spatiotemporal data becomes more complex, utilizing geospatial modeling on the Google Earth Engine (GEE) platform presents challenges in improving coding efficiency for experts and enhancing the coding capabilities of interdisciplinary users. To address these, we propose a framework for constructing a geospatial operator knowledge base tailored to the GEE JavaScript API. The framework includes an operator syntax knowledge table, an operator relationship frequency table, an operator frequent pattern knowledge table, and an operator relationship chain knowledge table. Using Abstract Syntax Tree (AST) techniques and frequent itemset mining, we extract operator knowledge from 185,236 real GEE scripts and syntax documentation, forming a structured knowledge base. Experimental results show the framework achieves over 90% accuracy, recall, and F1 score in operator knowledge extraction. When integrated with the Retrieval-Augmented Generation (RAG) strategy for LLM-based geospatial code generation tasks, the knowledge base enhances performance by 20-30%. Ablation studies further validate the importance of each knowledge table. This work advances geospatial code modeling techniques and supports the application of LLMs in geoinformatics, contributing to the integration of generative AI in the field.

## 1. Introduction

The rapid growth of spatiotemporal data has made geospatial modeling a crucial tool for uncovering the dynamic patterns of geographic phenomena(Breunig et al., 2020). However, the scale and complexity of such data present significant challenges. These datasets often adopt specialized formats (e.g., GeoJSON, GeoTIFF), and their analysis heavily relies on high-precision remote sensing imagery and substantial computational power. Traditional computing platforms struggle to efficiently store, process, and manage these data(Wang et al., 2018). Moreover, the pronounced spatial distribution characteristics of spatiotemporal data, along with the dependence on real-time processing and visualization during modeling, render conventional programming environments inadequate for meeting these demands(He et al., 2019).

To address these challenges, Google Earth Engine (GEE) has emerged as a cloud-based computing platform tailored for satellite imagery and spatiotemporal data(Tamiminia et al., 2020). GEE offers a wealth of built-in resources, including vast spatiotemporal datasets and powerful cloud computing capabilities. Users can efficiently access these resources by writing geospatial code, enabling complex geospatial modeling(Yang et al., 2022). Geospatial code is structured around "operators," which are fundamental units containing only functional functions, excluding variables or values(Granell et al., 2010). In other words, operators represent core actions within the code that execute specific functions, typically manifested as function calls that encapsulate particular tasks through standardized syntax. These operators are used for data processing, analysis, and visualization, providing computational, data manipulation, or logical operations(Garcia et al., 2007). Compared to general-purpose code, geospatial code differs significantly in operator naming, function design, and combination strategies(Hou et al., 2024a). While these differences enhance specificity and applicability, they also increase the complexity and barrier to entry for writing code, especially for non-expert users, thus limiting broader adoption.

The code generation capabilities of large language model (LLM), also known as NL2Code, offer a promising solution to the challenges mentioned above(Jiang et al., 2024). By expressing requirements in natural language, users can automatically generate code. However, existing LLMs are primarily trained on general-purpose code and lack specialized learning for geospatial code(Gu et al., 2024). As a result, the generated code often contains syntax or logical errors, or even exhibits "coding hallucinations," such as misspelled operator names, improper operator combinations, or the creation of non-existent operators(Hou et al., 2024b). Additionally, the complex structure of geospatial scripts makes operator knowledge difficult to leverage effectively, further hindering the performance of models in geospatial code generation tasks(Mansourian and Oucheikh, 2024).

To address the aforementioned issues, constructing a domain-specific knowledge base that contains rich operator knowledge presents a feasible solution(Wang et al., 2023). In domain knowledge injection methods, traditional domain-specific fine-tuning can enhance a model's specialized capabilities; however, it requires substantial computational resources, making it costly and difficult to scale(Jeong, 2024). Moreover, the effectiveness of this approach is heavily influenced by the quality and proportion of the training corpus, which can lead to knowledge dilution or weakened capabilities(Huang et al., 2023). In contrast, Retrieval-Augmented Generation (RAG), by constructing a specialized knowledge base and incorporating prompt engineering techniques, enables dynamic knowledge retrieval during the model generation process. This approach can significantly enhance the model's domain adaptability without the need for large-scale computational power, offering a more cost-effective solution(Gao et al., 2023).

Leveraging the extensive user base, rich open-source script resources, and diverse code logic and scales of the GEE platform(Liang et al., 2023), we propose a framework for constructing an operator knowledge base, as shown in Figure 1. The framework consists of three main steps: Collection and Organization, Statistical Analysis and Construction, and Validation and Evaluation. Using the GEE platform's JavaScript script database as the experimental dataset, we instantiated the GEE-OPs knowledge base from 185,236 real user scripts. The knowledge base includes an operator syntax knowledge table, an operator relationship frequency table, an operator frequent pattern knowledge table, and an operator relationship chain knowledge table. Evaluation metrics and a validation set were designed to systematically assess the accuracy and effectiveness of the knowledge base in LLM-based geospatial code generation tasks. The results demonstrate that the accuracy of the constructed knowledge base exceeds 90% across various metrics, with an average performance improvement of 20%-30% in multiple LLM geospatial code generation tasks. Additionally, ablation experiments further confirm the necessity of each knowledge table within the knowledge base. This work provides valuable insights, evaluation frameworks, and research directions for future knowledge base construction in LLM-based geospatial code generation tasks.

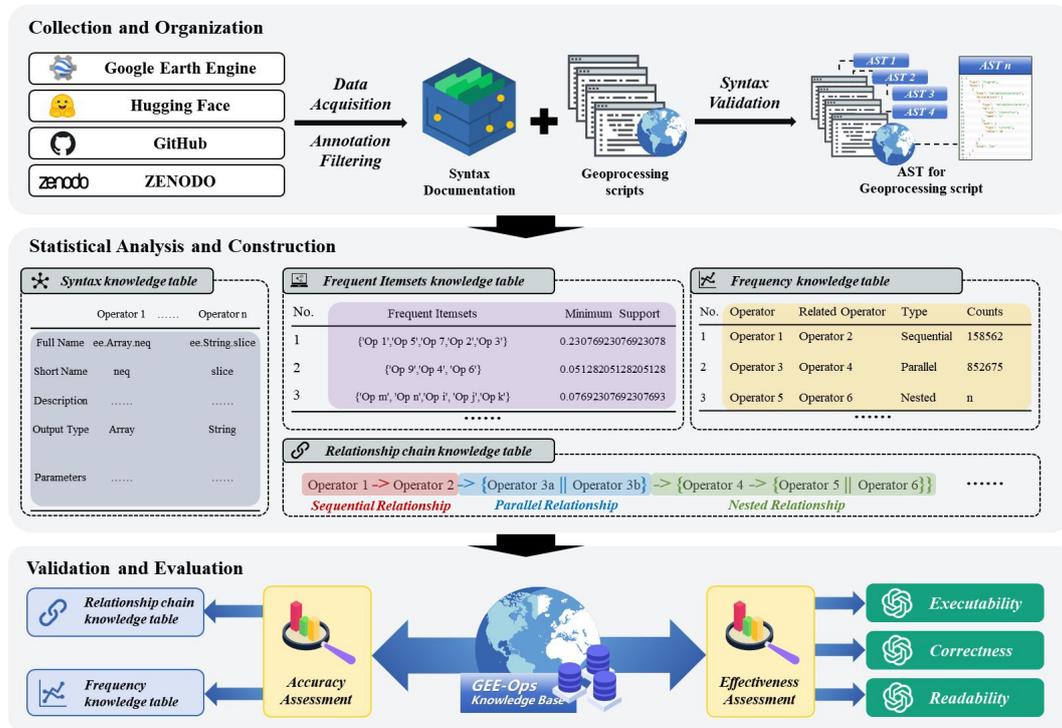

Fig.1 Framework for Constructing the GEE-OPs Operator Knowledge Base

The contributions of this paper are as follows:

- A framework for constructing a geospatial script operator relationship knowledge base is proposed.

- The GEE-OPs knowledge base is instantiated using real JavaScript scripts from the GEE platform.

- A systematic evaluation of the knowledge base's construction accuracy and its effectiveness in geospatial code generation tasks is conducted.

The structure of this paper is as follows:Chapter 2 provides a review of related research on three key topics: vertical domain applications of LLMs, code generation techniques, and knowledge base construction.Chapter 3 presents a detailed description of the methods for building the knowledge base and the corresponding implementation results.Chapter 4 offers a systematic quantitative evaluation and analysis of the knowledge base's accuracy and effectiveness based on construction metrics.Chapter 5 concludes the paper with a summary of the research findings and discusses potential directions for future work.

## 2. Related Work

## 2.1. Vertical Applications of LLMs

LLMs based on Transformer architecture and self-attention mechanisms have advanced significantly, demonstrating exceptional performance in general-domain tasks(Zhao et al., 2023). However, their effectiveness in vertical domains heavily depends on the availability and quality of specialized training data(Li et al., 2024a). Challenges include the substantial computational resources required for retraining and the risk of "knowledge hallucination," where models generate plausible but incorrect outputs due to the scarcity of domain-specific data in general-purpose corpora(Andriopoulos and Pouwelse, 2023). To address these challenges, research has focused on enhancing LLMs through domain specialization (LLM+X), with successful applications in fields such as law(Lai et al., 2024), finance(Wu et al., 2023), and biomedicine(Ling et al., 2023; Lu et al., 2024). However, in the GIS domain, while LLMs have been explored for tasks like knowledge question answering(Gupta et al., 2024; Zhang et al., 2024), knowledge extraction(Hou et al., 2024d; Hu et al., 2023), and spatiotemporal reasoning(Jin et al., 2023; Li et al., 2024b), geospatial code generation—a critical capability—remains largely unstudied.

## 2.2. Code Generation

The task of code generation (NL2Code) translates natural language requirements into source code(Zan et al., 2022). While early approaches relying on heuristic rules and probabilistic grammar frameworks were limited in handling complex requirements(Løkketangen and Olsson, 2010; Nymeyer and Katoen, 1997; Raychev et al., 2016), recent advances in LLMs trained on multimodal corpora—including text, web content, and code—have significantly improved performance. These models leverage enhanced context understanding, logical reasoning, and code generation capabilities, paving the way for automated programming(Jiang et al., 2024). However, due to the limited proportion of code data in training corpora, LLMs often produce code with syntax or logical errors, or generate non-executable yet superficially plausible outputs—a phenomenon known as "coding hallucination(Hou et al., 2024c)." This issue is particularly pronounced in geospatial code generation, where domain knowledge gaps result in operator name misspellings, improper operator combinations, and non-existent operators(Gramacki et al., 2024). Existing evaluation results show that the accuracy of geospatial code, from simple unit tests to complex task-based evaluations, typically falls below 20%(Gramacki et al., 2024; Hou et al., 2024b). Despite advancements in general programming tasks, no dedicated research has yet focused on optimizing LLMs for geospatial code generation.

## 2.3. Knowledge Bases for LLMs

Enhancing LLMs typically involves either improving learning capabilities through fine-tuning or injecting domain-specific knowledge via external knowledge bases(Ovadia et al., 2023). Compared to the resource-intensive and less flexible process of fine-tuning, knowledge injection offers a more efficient solution, leveraging high-quality knowledge bases to enhance domain adaptability(Czekalski and Watson, 2024). Constructing these knowledge bases typically involves collecting, processing, and organizing large-scale data from open-source platforms, supplemented by expert input for structural representation, thus optimizing task-specific

performance(Wan et al., 2024).Successful applications of knowledge bases have been demonstrated in domains such as biomedicine(Sung et al., 2021), law(Zhou et al., 2024), and finance(Zhao et al., 2024), supporting tasks like question answering, reasoning, and text generation. Among various methods, RAG dynamically retrieves relevant knowledge from a database to improve model generation(Gao et al., 2023). This approach significantly reduces resource consumption, ensures knowledge currency, and offers greater flexibility compared to fine-tuning.However, both general and geospatial code generation lack effective methods for representing and extracting operator relationships. While techniques like Abstract Syntax Trees (AST)(Noonan, 1985), Control Flow Graphs (CFG)(Gold, 2010), and Program Dependency Graphs (PDG)(Ferrante et al., 1987) are widely used in software engineering to analyze and classify source code, they focus primarily on syntax and structure rather than extracting operator relationship knowledge. The systematic development of operator relationship knowledge remains an unexplored and critical research direction.

## 3. Method

The construction of GEE-OPs involves three key stages: Collection and Organization, Statistical Analysis and Construction, and Validation and Evaluation。

### 3.1. Collection and Organization

### 3.1.1. Data Acquisition

Data acquisition encompasses high-quality geospatial scripts and official operator syntax knowledge. In this study, a large volume of user-contributed GEE scripts was collected from open-source platforms such as GitHub, Zenodo, and HuggingFace. A systematic search strategy, using keywords like "Google Earth Engine (GEE)", "JavaScript", "GIS", and "Geospatial", was employed, along with API integration, to automate the download of scripts. During the data collection process, special care was taken to ensure compliance with open-source licensing agreements and mitigate potential copyright risks. Scripts with specific licensing requirements were thoroughly documented, and compliance checks were conducted before use. Additionally, the official GEE platform documentation, which provides detailed descriptions of operator syntax and usage rules, was reviewed and compiled by domain experts to support the construction of the knowledge base. A total of 295,943 GEE scripts were collected, all written in JavaScript, spanning from September 2015 to September 2024. The script sizes ranged from 1 KB to 533 KB, with a total dataset size of 0.799 GB. The operator syntax data comprises 1,374 entries. The specific characteristics of the dataset are summarized in Table 1.

Tab. 1. Characteristics of the Collected GEE Script Dataset

| Attribute | Description |
| --- | --- |
| Programming Languages | JavaScript |
| Time Span | Sep.2015 - Sep.2024 |
| Quantity of Scripts | 295,943 |

| Attribute | Description |
| --- | --- |
| Script Size Range | 1kb-533kb |
| Total Dataset Size | 5.45 GB |
| Quantity of operator syntax knowledge | 1374 |

**3.1.2. Comment Filtering**

Due to variations in coding habits, programming styles, and language features among script developers, the collected geospatial scripts often contain redundant information such as comment text, debugging code, and outdated links. These extraneous elements can interfere with the extraction of operator relationships. To address this, a character-based rule constraint was applied to clean the scripts by removing all non-executable code segments and excess comments, while retaining the core functional code. This process optimized the script structure, reduced the dataset size, and facilitated subsequent structured representation based on ASTs.

**3.1.3. Syntax Checking**

The syntax correctness of open-source scripts cannot be fully guaranteed, and syntax errors may introduce incorrect knowledge during the extraction of operator relationships. To address this issue, AST technology was employed to structurally represent and validate the syntax of geospatial scripts(Noonan, 1985). AST parsing transforms the scripts into a tree structure, represented in JSON format, with each node corresponding to an element in the source code (such as operators, variables, and parameters). This representation allows for precise differentiation of operators from other code elements using "key names" and clarifies operator dependencies and execution order through parentheses hierarchy and indentation relationships. Additionally, scripts with syntax errors will fail to convert during the AST parsing process, thereby being automatically filtered out during the syntax validation phase, preventing erroneous information from entering the knowledge base. This method eliminates the need for manual verification, significantly improving processing efficiency while enhancing the accuracy and reliability of operator relationship extraction. After syntax checking, 234,067 syntactically correct scripts were retained.

**3.2. Statistical Analysis and Construction**

The GEE-OPs knowledge base consists of four components: the operator syntax knowledge table, operator relationship frequency knowledge table, operator frequent itemset knowledge table, and operator relationship chain knowledge table. The operator syntax knowledge table provides accurate information on operator names and their usage syntax. The operator relationship frequency knowledge table records the types and frequencies of operator pair relationships across a large dataset of scripts. The operator frequent itemset knowledge table, derived from the operator relationship frequency table, uses frequent itemset mining techniques to extract high-frequency operator combinations. The operator relationship chain knowledge table builds upon the operator relationship frequency table and the operator frequent itemset knowledge table, further supplementing the contextual information of the scripts. All knowledge tables are stored in CSV format. The specific design of each knowledge table is shown in Figure 2.

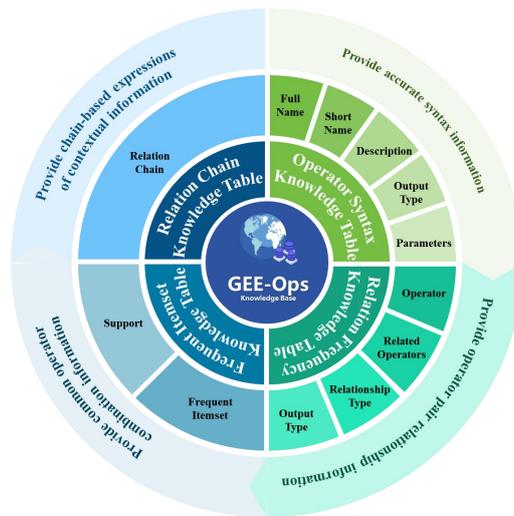

Fig.2 Design of the Knowledge Tables in the GEE-OPs Knowledge Base

### 3.2.1. Operator Syntax Knowledge Table

The operator syntax knowledge table is constructed based on 1,374 operator syntax data points collected from the official platform.

### 3.2.2. Operator Relationship Frequency Knowledge Table

Based on control flow and data flow theories in computer compiler principles(Beck et al., 1991), and in conjunction with the typical workflows and logical structures of geospatial data processing(Shi and Walford, 2012), operator relationships in geospatial scripts can be categorized into three main types: sequential relationships, parallel relationships, and nested relationships(Cao et al., 2021; Collins et al., 2006; Garani, 2003). Sequential relationships represent the most fundamental type of operator interaction, often occurring in operations such as variable assignment, arithmetic calculations, or function calls. For example, in surface data processing or meteorological model execution, a series of data transformations or computational steps typically need to be performed in a strict sequence to ensure the accuracy of the results. Parallel relationships, on the other hand, allow multiple operations to be executed simultaneously, without dependence on a specific order or the results of other operations. This type of relationship is commonly found in independent computational tasks, such as concurrently calculating statistical features across multiple regions or performing format conversions on multiple datasets in parallel. Nested relationships reflect the logical dependencies within multi-level control structures, such as nested loops or complex conditional statements. In multi-criteria remote sensing data analysis or complex spatial data query tasks, nested structures can significantly improve the efficiency of data processing and querying. Figure 3 illustrates geospatial code examples that demonstrate these operator relationships.

```
1    Sequential Relationship
2    var image = ee.Image('LANDSAT/LC08/C01/T1_SR/LC08_044034_20140318');
3    var ndvi = image.normalizedDifference(['B5', 'B4']);
4    Map.addLayer(ndvi, {min: 0, max: 1}, 'NDVI');
5
6    Parallel Relationship
7    var image1 = ee.Image('LANDSAT/LC08/C01/T1_SR/LC08_044034_20140318');
8    var image2 = ee.Image('LANDSAT/LC08/C01/T1_SR/LC08_044035_20140318');
9    var ndvi1 = image1.normalizedDifference(['B5', 'B4']);
10   var ndvi2 = image2.normalizedDifference(['B5', 'B4']);
11   Map.addLayer(ndvi1, {min: 0, max: 1}, 'NDVI1');
12   Map.addLayer(ndvi2, {min: 0, max: 1}, 'NDVI2');
13
14   Nested Relationship
15   var collection = ee.ImageCollection('LANDSAT/LC08/C01/T1_SR')
16     .filterBounds(ee.Geometry.Point(-122.262, 37.8719))
17     .map(function(image) {
18       var ndvi = image.normalizedDifference(['B5', 'B4']);
19       return ndvi;
20     });
21   var maxNDVI = collection.qualityMosaic('nd');
22   Map.addLayer(maxNDVI, {min: 0, max: 1}, 'Max NDVI');
```

Fig.3 Geospatial code examples illustrating operator relationships

During the data collection and cleaning phase, we employed AST representations to structurally encode geospatial scripts that conform to syntactic rules. To further quantify the types of operator relationships and their frequencies within the scripts, it is necessary to identify specific structures in the AST that correspond to different operator relationships and develop appropriate statistical methods. In an AST, nodes represent the fundamental units of code logic, with each node corresponding to a statement or expression in the script. Since a single statement or expression may involve one or more operator calls (e.g., nested function calls or multi-function operations), each node contains a set of operators, recording all the functional calls within that node.First, we used a Depth-First Search (DFS) algorithm to traverse the AST(Tarjan, 1972), marking the depth of each node and assigning a primary label to each node. Within each node, all operators are given a secondary label based on their order of appearance in the statement, ensuring that each operator has a unique identifier. Operators within the same node are always in a sequential relationship, as the order of operator calls in a single statement or expression is fixed. For different nodes, if the output of an operator in node A is used by an operator in node B, a sequential relationship is marked between the last operator in node A and the first operator in node B. For sibling nodes with the same parent, the intersection of their operator sets is computed; if the intersection is empty, a parallel relationship is marked between the child nodes. Additionally, by recording the hierarchical structure of parent-child nodes, control flow nodes in the AST (e.g., IfStatement, ForStatement, and WhileStatement) are identified to delineate the control scope and influence of parent nodes on their child nodes, thereby marking nested relationships.The final results are recorded in the Operator Relationship Frequency Knowledge Table, which includes the frequency of each relationship type and the corresponding operator pairs, providing valuable support for subsequent geospatial code analysis and knowledge base construction.

### 3.2.3. Operator Frequent Itemset Knowledge Table

The operator relationship frequency knowledge table describes pairwise relationships between operators, revealing the operators that are more likely to follow after a given operator. This provides a foundation for identifying high-frequency operator combination patterns. By applying frequent itemset mining, commonly used operator combination patterns can be extracted, reflecting typical code statements or functional workflows in geospatial scripts. Based on the operator relationship frequency knowledge table, this study employs the classical FP-Growth algorithm to extract frequent operator patterns and generate a frequent operator relationship knowledge table.

FP-Growth is an efficient algorithm for frequent itemset mining, which avoids candidate generation and pruning operations by constructing and traversing a frequent pattern tree (FP-Tree), significantly improving memory efficiency and computational speed(Grahne and Zhu, 2005). Compared to the traditional Apriori algorithm, FP-Growth demonstrates superior performance in handling large-scale operator relationships, such as the dataset in this study comprising hundreds of thousands of records(Wang and Jiao, 2020). The key parameter of the algorithm is the minimum support threshold (min_support), defined as the proportion of the frequency of an operator combination to the total frequency of all operator relationships. In this study, the minimum support threshold is set to 0.05 to balance the trade-off between pattern coverage and computational efficiency(Mallik et al., 2018; Thurachon and Kreesuradej, 2021). To further enhance the performance and resource efficiency of FP-Growth, the following optimization strategies were adopted, tailored to the characteristics of the operator relationship frequency knowledge table: filtering low-frequency items prior to FP-Tree construction to reduce data size; implementing a batch processing strategy to partition the data for computation and reduce memory usage; utilizing sparse matrix representations to optimize data storage and processing efficiency; and dynamically adjusting the support threshold by starting with a higher threshold to identify significant patterns and progressively lowering it to capture sparse but critical patterns.

### 3.2.4. Operator Relationship Chain Knowledge Table

The Operator Relationship Frequency Knowledge Table and Frequent Itemset Knowledge Table primarily describe the overall frequency and patterns of operator relationships, but they lack specific context information about the individual scripts. To address this gap, we utilize the hierarchical structure of AST to accurately capture the control flow and data flow characteristics of the code, establishing a symbolic system. Based on this system, we further extract the operator relationship chains for each script, using a chain structure to succinctly represent the execution order and dependencies of operators in the specific script. Additionally, the design of the symbolic system preserves semantic information, such as execution logic, through specific symbols. The symbolic system consists of three expression structures, each designed to accurately describe the three types of relationships within an operator chain. The detailed design of the symbolic system is shown in Table 2. The method of traversing and constructing this symbolic system aligns with the process used to build the Operator Relationship Frequency Knowledge Table.

Tab.2 Design of the Symbolic System for Operator Relationship Chains

| Relationship Type | Abstract Symbol | Example |
|:---:|:---:|:---:|
| Sequential | -> | Operator 1 -> Operator 2 |
| Parallel | \|\| | > {Operator 3a \|\| Operator 3b} |

| Relationship Type | Abstract Symbol | Example |
|---|---|---|
| Nested | -> {...} | -> {Operator 4 -> {Operator 5 \|\| Operator 6}} |

### 3.3. Construction Results

The Operator Syntax Knowledge Table comprises five core attributes: *full_name*, *short_name*, *description*, *output_type*, and *parameters*, containing a total of 1,374 records (Figure 4-a). The Operator Relationship Frequency Knowledge Table includes four attributes: *operator*, *related operator*, *relationship type*, and *relationship frequency*, documenting 8,074 operator relationships with frequency information (Figure 4-b). The Operator Frequent Itemset Knowledge Table, generated through frequent itemset mining, contains attributes such as *antecedent*, *consequent*, *antecedent support*, *consequent support*, *support*, *confidence*, and *lift*, with 52,851 frequent patterns recorded (Figure 4-c). The Operator Relationship Chain Knowledge Table consists of two components: *script name* and *relationship chain extraction result*, where each script corresponds to one extracted relationship chain, totaling 211,659 records (Figure 4-d). The related code and construction results are publicly available at https://figshare.com/s/6d42f6335f3f6254ea14.

*(a) Operator Syntax Knowledge Table*

*(b) Operator relationship frequency knowledge table*

*(c) Operator relationships frequent pattern knowledge table*

*(d) Operator Relationship Chain Knowledge Table*

Fig.4 Structure and Sample Entries of the GEE-OPs Knowledge Tables

## 4. Evaluation

The evaluation is conducted in two aspects: (1) assessing the accuracy of the GEE-OPs knowledge base construction results, and (2) evaluating the effectiveness of the GEE-OPs knowledge base in improving LLMs' geospatial code generation capabilities.

**4.1. Accuracy Evaluation**

The GEE-OPs knowledge base consists of four knowledge tables. Among them, the Operator Syntax Knowledge Table is constructed based on objective data collection, and the Operator Frequent Itemset Knowledge Table is generated through frequent itemset mining from the Operator Relationship Frequency Knowledge Table, both of which yield unique and deterministic results, thus requiring no further accuracy evaluation. The evaluation focus is therefore placed on the Operator Relationship Frequency Knowledge Table and the Operator Relationship Chain Knowledge Table.Given the lack of existing research methods for direct comparison on operator relationship extraction in scripts, this study adopts an expert-based ground truth annotation approach. Using the quartile distribution of operator scale in scripts, 30 representative scripts are selected for quantitative evaluation: 10 small-scale scripts (involving no more than 20 operator relationships), 10 medium-scale scripts (20-80 operator relationships), and 10 large-scale scripts (more than 80 operator relationships).

**4.1.1. Evaluation of the Operator Relationship Frequency Knowledge Table**

The evaluation of the Operator Relationship Frequency Knowledge Table focuses on accurately identifying the types of operator relationships occurring in scripts. First, domain experts manually annotate the actual operator relationship types in each script to generate the ground truth. Subsequently, each operator relationship (sample) extracted using the proposed method is categorized as follows:

- **TP (True Positive):** Correctly identified and actually existing operator relationships.
- **FP (False Positive):** Identified operator relationships that do not actually exist.
- **FN (False Negative):** Operator relationships that exist but were not identified.

The following metrics are employed to quantitatively evaluate the samples:

- **Accuracy:** Proportion of correctly identified operator relationships among all identified relationships, representing the overall correctness of the method. The formula is:

$$\text{Accuracy} = \frac{TP}{TP + FP + FN} \#(1)$$

- **Recall:** Proportion of correctly identified operator relationships among all actual operator relationships. High recall indicates the method effectively avoids missing important relationships and captures most true relationships. The formula is:

$$\text{Recall} = \frac{TP}{TP + FN} \#(2)$$

- **Precision:** Proportion of correctly identified operator relationships among all identified

relationships, reflecting the reliability of the method. Higher precision indicates fewer false identifications. The formula is:

$$\text{Precision} = \frac{TP}{TP + FP} \#(3)$$

- **F1 Score:** The harmonic mean of precision and recall, balancing the trade-off between capturing true relationships and reducing false identifications. The formula is:

$$F1 = 2 \times \frac{\text{Precision} \times \text{Recall}}{\text{Precision} + \text{Recall}} \#(4)$$

- **Coefficient of Variation (CV):** The ratio of the standard deviation to the mean of the metrics, measuring the stability of the evaluation indicators across samples. The formula is:

$$CV = \frac{\sigma}{\mu} \#(5)$$

The evaluation results are presented in Table 4. For small-scale scripts, GEE-OPs accurately predicted nearly all operator relationships, achieving perfect predictions in 4 scripts and only a single misclassification in 6 scripts, demonstrating exceptionally high accuracy. For medium-scale and large-scale scripts, although the misclassification rate increased slightly, accuracy remained high, with an average accuracy exceeding 0.73. Overall, the average accuracy across all scripts reached 0.87, while recall, precision, and F1 score all achieved 0.93, indicating that GEE-OPs effectively extracts operator relationship types and frequencies from scripts. Additionally, the coefficient of variation (CV) for all metrics was below 0.1, reflecting minimal fluctuation and high stability in the evaluation results.

Tab. 4 Evaluation Results of the Operator Relationship Frequency Knowledge Table

| Script_name | TP | FP | FN | Accuracy | Recall | Precision | F1 Score |
|---|---|---|---|---|---|---|---|
| **Small-scale Scripts** | | | | | | | |
| Script_01 | 8 | 0 | 0 | 1.00 | 1.00 | 1.00 | 1.00 |
| Script_02 | 5 | 1 | 0 | 0.83 | 1.00 | 0.83 | 0.91 |
| Script_03 | 6 | 0 | 0 | 1.00 | 1.00 | 1.00 | 1.00 |
| Script_04 | 3 | 0 | 0 | 1.00 | 1.00 | 1.00 | 1.00 |
| Script_05 | 9 | 0 | 0 | 1.00 | 1.00 | 1.00 | 1.00 |
| Script_06 | 4 | 0 | 1 | 0.80 | 0.80 | 1.00 | 0.89 |
| Script_07 | 5 | 1 | 0 | 0.83 | 1.00 | 0.83 | 0.91 |
| Script_08 | 7 | 1 | 0 | 0.88 | 1.00 | 0.88 | 0.93 |
| Script_09 | 4 | 0 | 1 | 0.80 | 0.80 | 1.00 | 0.89 |
| Script_10 | 18 | 1 | 0 | 0.95 | 1.00 | 0.95 | 0.97 |

| Script_name | TP | FP | FN | Accuracy | Recall | Precision | F1 Score |
|---|---|---|---|---|---|---|---|
| **Medium-scale Scripts** | | | | | | | |
| Script_11 | 55 | 3 | 14 | 0.76 | 0.80 | 0.95 | 0.87 |
| Script_12 | 48 | 1 | 2 | 0.94 | 0.96 | 0.98 | 0.97 |
| Script_13 | 32 | 2 | 3 | 0.86 | 0.91 | 0.94 | 0.93 |
| Script_14 | 26 | 0 | 8 | 0.76 | 0.76 | **1.00** | 0.87 |
| Script_15 | 13 | 3 | 0 | 0.81 | **1.00** | **0.81** | 0.90 |
| Script_16 | 37 | 5 | 3 | 0.82 | 0.93 | 0.88 | 0.90 |
| Script_17 | 44 | 8 | 5 | 0.77 | 0.90 | 0.85 | 0.87 |
| Script_18 | 29 | 3 | 8 | **0.73** | **0.78** | 0.91 | **0.84** |
| Script_19 | 38 | 5 | 2 | 0.84 | 0.95 | 0.88 | 0.92 |
| Script_20 | 28 | 0 | 1 | 0.97 | 0.97 | 1.00 | 0.98 |
| **Large-scale Scripts** | | | | | | | |
| Script_21 | 87 | 9 | 5 | 0.86 | 0.95 | 0.91 | 0.93 |
| Script_22 | 384 | 13 | 15 | 0.93 | 0.96 | 0.97 | 0.96 |
| Script_23 | 692 | 48 | 4 | 0.93 | 0.99 | 0.94 | 0.96 |
| Script_24 | 61 | 9 | 13 | **0.73** | 0.82 | 0.87 | 0.85 |
| Script_25 | 473 | 27 | 5 | 0.94 | 0.99 | 0.95 | 0.97 |
| Script_26 | 896 | 38 | 19 | 0.94 | 0.98 | 0.96 | 0.97 |
| Script_27 | 268 | 48 | 2 | 0.84 | 0.99 | 0.85 | 0.91 |
| Script_28 | 421 | 17 | 15 | 0.93 | 0.97 | 0.96 | 0.96 |
| Script_29 | 319 | 32 | 16 | 0.87 | 0.95 | 0.91 | 0.93 |
| Script_30 | 136 | 9 | 35 | 0.76 | 0.80 | 0.94 | 0.86 |
| **Coefficient of Variation** | | | | 0.10 | 0.09 | 0.06 | 0.05 |
| **Average value** | | | | 0.87 | 0.93 | 0.93 | 0.93 |

**4.1.2. Evaluation of the Operator Relationship Chain Knowledge Table**

The evaluation of the Operator Relationship Chain Knowledge Table focuses on the accuracy of its structural representation and semantic information. The chain structure is designed to clearly reflect the execution order, parallel operations, and nested relationships in the script, ensuring that the dependencies between operators are fully and accurately expressed. The semantic information retains operator names and their combinations, accurately reflecting the functional calls and contextual associations of each operator. To evaluate the structural features of the chain, Longest Common Subsequence (LCS) and N-gram Similarity metrics are used(Abdeljaber, 2021; Nakatsu et al., 1982). For semantic features, Siamese Similarity and BERT Similarity metrics are applied(Ma et al., 2022; Mrinalini et al., 2022):

- **Longest Common Subsequence (LCS):** Measures the length of the longest subsequence common to two sequences while maintaining order. It evaluates the matching degree of operator chains in terms of execution order, particularly useful for analyzing the correctness of order-sensitive scripts.

- **N-gram Similarity:** Compares fixed-size N-grams (adjacent character combinations) between two strings to capture local co-occurrence patterns in operator chains and identify common structures or repeated patterns.

- **Siamese Similarity:** Uses a deep learning model to generate embedding vectors and optimizes semantic distance between inputs, allowing precise evaluation of the semantic similarity of operator chains in specific domains.

- **BERT Similarity:** Employs the BERT model to generate deep semantic embeddings and calculates cosine similarity to assess the semantic matching capability of operator chains.

The evaluation results are presented in Table 5. Overall, the semantic similarity metrics (Siamese and BERT) and structural similarity metrics (LCS and N-gram) fall within the range of 0.79 to 0.89, with coefficients of variation (CV) between 0.07 and 0.15, indicating high accuracy and stability in the results. Among these, the CV for N-gram is 0.15, slightly higher than the others, likely due to its sensitivity to local features. In contrast, the CV for BERT is only 0.07, reflecting its robustness in handling complex semantic relationships.

Tab. 5 Evaluation Results of Operator Relationship Chain Knowledge Table

| Script_name | LCS | Ngram | Siamese | BERT |
|---|---|---|---|---|
| **Small-scale Scripts** | | | | |
| Script_01 | 0.88 | 0.76 | 0.78 | 0.96 |
| Script_02 | 0.75 | 0.71 | 0.93 | 0.98 |
| Script_03 | 0.72 | 0.93 | 0.94 | 0.86 |
| Script_04 | 0.98 | 0.74 | 0.89 | 0.82 |
| Script_05 | 0.99 | 0.71 | 0.94 | 0.85 |
| Script_06 | 0.94 | 0.82 | 0.87 | 0.89 |
| Script_07 | 0.79 | 0.66 | 0.88 | 0.96 |
| Script_08 | 0.73 | 0.92 | 0.86 | 0.97 |

| Script_name | LCS | Ngram | Siamese | BERT |
|---|---|---|---|---|
| **Script_09** | 0.91 | 0.63 | 0.76 | 0.8 |
| **Script_10** | 0.83 | 0.99 | 0.78 | 0.9 |
| **Medium-scale Scripts** | | | | |
| **Script_11** | 0.74 | 0.91 | 0.76 | 0.88 |
| **Script_12** | 0.85 | 0.68 | 0.91 | 0.84 |
| **Script_13** | 0.71 | 0.6 | 0.83 | 0.82 |
| **Script_14** | 0.97 | 0.93 | 0.88 | 0.87 |
| **Script_15** | 0.78 | 0.88 | 0.98 | 0.99 |
| **Script_16** | 0.9 | 0.89 | 0.81 | 0.86 |
| **Script_17** | 0.79 | 0.91 | 0.85 | 0.9 |
| **Script_18** | 0.86 | 0.63 | 0.94 | 0.94 |
| **Script_19** | 0.86 | 0.74 | 0.81 | 0.87 |
| **Script_20** | 0.76 | 0.65 | 0.77 | 0.99 |
| **Large-scale Scripts** | | | | |
| **Script_21** | 0.99 | 0.95 | 0.82 | 0.99 |
| **Script_22** | 0.93 | 0.85 | 0.79 | 0.85 |
| **Script_23** | 0.98 | 0.73 | 0.98 | 0.9 |
| **Script_24** | 0.97 | 0.63 | 0.95 | 0.86 |
| **Script_25** | 0.88 | 0.72 | 0.91 | 0.86 |
| **Script_26** | 0.98 | 0.73 | 0.97 | 0.81 |
| **Script_27** | 0.73 | 0.89 | 0.95 | 0.92 |
| **Script_28** | 0.76 | 0.86 | 0.8 | 0.9 |
| **Script_29** | 0.71 | 0.95 | 0.97 | 0.81 |
| **Script_30** | 0.8 | 0.79 | 0.88 | 0.86 |
| **Coefficient of Variation** | **0.12** | **0.15** | **0.08** | **0.07** |
| **Average value** | **0.85** | **0.79** | **0.87** | **0.89** |

### 4.2. Effectiveness Evaluation

The effectiveness evaluation focuses on analyzing the improvement in code generation capabilities of mainstream LLMs after integrating GEE-OPs and applying the RAG framework. Detailed information about the selected models is provided in Table 6.

Tab. 6 Specifications of Selected LLMs

| Category | Model Name | Company | Size | Date |
|---|---|---|---|---|
| **Commercial LLMs** | GPT-4o-Mini | OpenAI | N/A | 2024 |
| **General LLMs** | LLaMA3 | Meta AI | 8B | 2024 |
| **Code Generation LLMs** | Code Llama | Meta AI | 13B | 2023 |

This study evaluates the geospatial code generation capabilities of LLMs using the GeoCode-Bench dataset, which comprises 20 representative tasks(Hou et al., 2024b). GeoCode-Bench includes a variety of geospatial problems such as spatial data processing, map generation, and geospatial analysis, offering a comprehensive assessment of the models' performance in geospatial code generation. The methodology is as follows: First, natural language queries are converted into vector representations using LLMs via the LangChain tool(Mahadevan and Raman, 2023). Concurrently, the knowledge base entries are transformed into vector representations using the SentenceTransformer model(Devika et al., 2021). To ensure compatibility between the natural language queries and knowledge base entries, both adopt BERT-based vector embeddings, allowing comparison and matching within the same semantic space. Next, FAISS (Facebook AI Similarity Search) is employed to perform nearest-neighbor matching between the query vectors and the knowledge base vectors(Krisnawati et al., 2024). FAISS uses cosine similarity as the metric and has been optimized to enhance retrieval efficiency and accuracy. The knowledge base is organized in CSV format, with each record containing precomputed vector representations and corresponding geospatial algorithms and solutions. These records undergo rigorous quality control and validation to ensure data comprehensiveness and reliability. The retrieval results are merged with the original queries to create new prompts, guiding the LLMs to generate the final code. The experiment compares the performance of prompt templates enhanced by knowledge base retrieval with zero-shot generation methods, evaluated using metrics such as code correctness, readability, and execution effectiveness(Pourpanah et al., 2022). The workflow for using the knowledge base is illustrated in Figure 5.

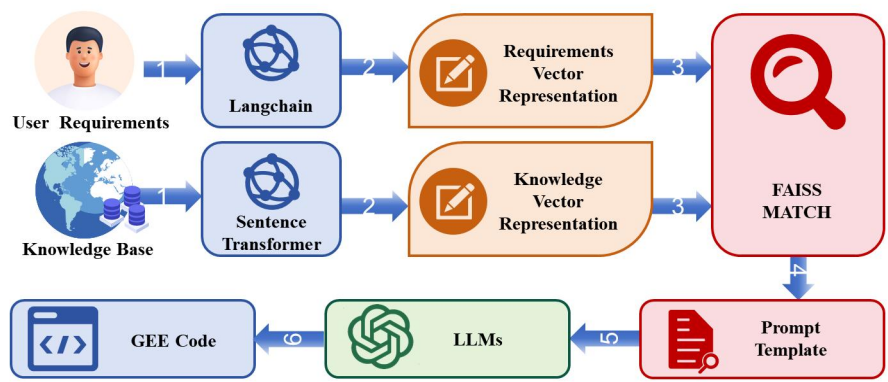

Fig.5 Workflow of RAG for Geospatial Code Generation

To evaluate the generated geospatial code, three key metrics were employed: executability, execution correctness, and readability. Executability is defined as the ability of the generated code to successfully run in the target environment, with a score awarded upon successful execution. Execution correctness refers to whether the output of the code meets the analytical objectives of the task. For open-ended requirements, correctness is determined based on expert judgment, with the code considered correct as long as it executes properly and achieves the intended goals. Readability is assessed through blind scoring by five experts, with a maximum score of 5 points. The final score for each task is calculated by averaging the scores after removing the highest and lowest ratings. The average score across all tasks within the same group and model is taken as the model's final performance score for the comparison or ablation study. All metric scores are normalized to a percentage scale, rounded to one decimal place.

The evaluation aims to address two key questions:

- EQ1: Is the approach effective across various LLMs?
- EQ2: Can the necessity of the four knowledge tables be validated through ablation experiments on GPT-4o-mini?

### 4.2.1. EQ1

The evaluation results are presented in Table 7. The performance of various models in the zero-shot setting was relatively modest. For executability and execution correctness, the metrics ranged from 0.15 to 0.47, with none exceeding 0.5. However, after incorporating the GEE-OPs knowledge base for supplementary knowledge, performance improved significantly, with increases ranging from 0.26 to 0.47. The optimized metrics ranged from 0.55 to 0.85, demonstrating a substantial enhancement in model reliability.In contrast, readability showed only minor improvements, likely due to the GEE-OPs knowledge base being focused solely on operator-related knowledge and not addressing semantic expressions such as comments. This limitation highlights an important direction for future optimization. Overall, the average performance improvement across the three metrics ranged from 0.24 to 0.32, underscoring the effectiveness of GEE-OPs in enhancing geospatial code generation capabilities across various LLMs.

Tab. 7 Evaluation Results of EQ1

| LLMs | Indicators | Executability | Correctness | Readability | Average |
|---|---|---|---|---|---|
| GPT-4o-Mini | zero-shot | 0.47 | 0.45 | 0.53 | 0.48 |
|  | RAG | 0.85 | 0.71 | 0.60 | 0.72 |
|  | **Difference** | **+0.38** | **+0.26** | **+0.07** | **+0.24** |
| LLaMA3-8B | zero-shot | 0.24 | 0.15 | 0.47 | 0.29 |
|  | RAG | 0.68 | 0.55 | 0.52 | 0.58 |
|  | **Difference** | **+0.44** | **+0.40** | **+0.06** | **+0.30** |

| LLMs | Indicators | Executability | Correctness | Readability | Average |
|---|---|---|---|---|---|
| Code Llama-13B | zero-shot | 0.28 | 0.17 | 0.45 | 0.30 |
| | RAG | 0.71 | 0.64 | 0.50 | 0.62 |
| | Difference | +0.43 | +0.47 | +0.06 | +0.32 |

### 4.2.2. EQ2

The evaluation results are presented in Table 8. Given that readability did not show significant improvement in the EQ1 evaluation, it is excluded as a metric in subsequent assessments. Additionally, as the Operator Syntax Knowledge Table plays a foundational role in operator syntax knowledge evaluation, it is enabled by default in all evaluations of knowledge table combination strategies. Independent effects of other knowledge tables are not assessed separately. The evaluation results indicate that as the knowledge tables are incrementally incorporated, GPT-4o-mini exhibits substantial improvements across all metrics in geospatial code generation capabilities. This underscores the necessity of each knowledge table in enhancing the geospatial code generation performance of LLMs.

Tab. 8 Evaluation Results of EQ2

| Ta.1 | Ta.2 | Ta.3 | Ta.4 | Executability | Correctness | Average |
|---|---|---|---|---|---|---|
| × | × | × | × | 0.47 | 0.45 | 0.46 |
| √ | × | × | × | 0.55(+0.08) | 0.50(+0.05) | 0.53(+0.07) |
| √ | √ | × | × | 0.68(+0.21) | 0.60(+0.15) | 0.64(+0.18) |
| √ | × | √ | × | 0.65(+0.18) | 0.58(+0.13) | 0.62(+0.16) |
| √ | × | × | √ | 0.63(+0.16) | 0.57(+0.12) | 0.60(+0.14) |
| √ | √ | √ | × | 0.75(+0.28) | 0.68(+0.23) | 0.72(+0.26) |
| √ | √ | × | √ | 0.78(+0.31) | 0.70(+0.25) | 0.74(+0.28) |
| √ | × | √ | √ | 0.77(+0.30) | 0.69(+0.24) | 0.73(+0.27) |
| √ | √ | √ | √ | **0.85(+0.38)** | **0.71(+0.26)** | **0.78(+0.32)** |

## 5. Conclusion

This study introduces and constructs the GEE-OPs geospatial script operator knowledge base, developed on the GEE platform, to provide a systematic solution to the critical challenges faced by large language models (LLMs) in geospatial code generation. By organizing domain-specific knowledge into four key components—operator syntax, relationship frequency, frequent itemsets, and relationship chains—GEE-OPs enables efficient knowledge representation and retrieval. In operator relationship extraction tasks, the knowledge base demonstrates outstanding performance, with all core metrics (accuracy, recall, precision, and F1 score) exceeding 90%, affirming its high precision in knowledge extraction. In geospatial code generation tasks, GEE-OPs enhances LLM performance by 20%-30%, significantly improving the models' applicability and reliability. Ablation experiments further confirm the indispensable role of each knowledge base component in advancing model performance.

### 5.1. Advantages and Significance

This study addresses the critical gap in applying large language models (LLMs) to geospatial code generation, specifically the lack of domain-specific knowledge on operator syntax, relationships, and combinations. It introduces a novel method for extracting operator relationships in geospatial code, systematically bridging gaps in the literature. By constructing the GEE-OPs knowledge base, this research offers a robust solution for operator knowledge extraction, significantly enhancing the accuracy and reliability of geospatial code generation.

In the context of geographic information science (GIScience), this study presents an innovative framework for geospatial code generation, integrating the GEE-OPs knowledge base with a Retrieval-Augmented Generation (RAG) framework. This approach not only improves LLM performance but also provides a practical foundation for automated code generation, data processing, and spatial analysis, supporting GIScience research.

For readers, this study highlights the transformative potential of LLMs in geospatial code generation by lowering technical barriers. Domain experts can efficiently generate reliable geospatial code without model fine-tuning, while interdisciplinary users can produce accurate code tailored to their needs, fostering cross-disciplinary analysis. Beyond enhancing existing model performance, the GEE-OPs knowledge base offers a valuable resource for advancing geospatial programming knowledge and broadening LLM applications in GIScience, remote sensing, and spatial data processing, promoting integration with other disciplines.

### 5.2. Limitations and Future Work

This study focuses on extracting operator syntax and relationships from a large set of real-world scripts, laying the foundation for knowledge base research in geospatial code generation. However, while the current knowledge base emphasizes operator syntax and relationships, it has yet to fully incorporate other contextual information that could enhance the accuracy and applicability of the generated code. For example, integrating real-time variable inputs and user-specific parameter data could further optimize model performance, though this would require additional data sources and more complex processing methods. Despite the significant improvements GEE-OPs has demonstrated in geospatial code generation tasks, its applicability to more specialized or emerging tasks still requires further investigation. Beyond code generation, the knowledge base can be extended to tasks such as code completion, operator prediction, error correction, summary generation, and automatic annotation, providing comprehensive technical support for geospatial coding. These areas will be important directions for future research and exploration.

**Disclosure statement**

No potential conflict of interest was reported by the authors.

**Funding**

The work was supported by National Natural Science Foundation of China (no. 41930107).

**Data and codes availability statement**

The data and codes supporting the findings of this study are openly available at the following anonymous Figshare link: https://figshare.com/s/6d42f6335f3f6254ea14. The repository includes all scripts, datasets, and step-by-step instructions required to reproduce the results presented in the paper. Detailed documentation is provided to facilitate replication and ensure transparency.